\documentclass[sigplan,screen,natbib=false]{acmart}
\setcopyright{none}
\renewcommand\footnotetextcopyrightpermission[1]{} % removes footnote

\usepackage[backend=biber,style=numeric-comp,sorting=none]{biblatex}

\DeclareSourcemap{
    \maps[datatype=bibtex]{
    }
}
\AtEveryBibitem{
\clearfield{isbn}
\clearfield{editor}
\clearfield{address}
  \ifentrytype{inproceedings}{
    \clearfield{doi}
  }{}
  \ifentrytype{article}{
    \clearfield{doi}
  }{}
  \ifentrytype{book}{
    \clearfield{doi}
  }{}
  \ifentrytype{inbook}{
    \clearfield{doi}
  }{}
}
\AtBeginDocument{%
  }

\definecolor{dynamopurple}{RGB}{95, 85, 200}
\definecolor{dynamolightpurple}{HTML}{EFEEFA}
\definecolor{dynamodarkpurple}{HTML}{423A8A}

\definecolor{dynamoblue}{RGB}{3, 122, 204}
\definecolor{lapred}{RGB}{174, 0, 16}
\newcommand{\coloredname}[2]{%
  \tikz[baseline=(char.base)]{%
    \node[shape=rectangle,fill=#1,inner sep=1.5pt,text=white] (char) {\texttt{#2}};%
  }\xspace
}

\usepackage{hyperref}
\hypersetup{
    colorlinks=true,
    linkcolor=dynamopurple,
    filecolor=dynamopurple,      
    urlcolor=dynamopurple,
    citecolor=dynamopurple,
    }

\newcommand{\welike}{\coloredname{dynamopurple}{We like}}
\newcommand{\wehate}{\coloredname{dynamoblue}{We hate}}

\newcommand{\jiahui}[1]{\textbf{\color{dynamoblue}Jiahui: #1}}
\newcommand{\emmet}[1]{\textbf{\color{red}Emmet: #1}}

\usepackage{setspace}

\usepackage{tikz}

\usepackage{xspace}
\usepackage{subcaption}

\usepackage{fancyvrb}
\definecolor{highlightcolor}{HTML}{b06a07}

\newcounter{myctr}
\newcommand{\casestudy}[1]{%
  \stepcounter{myctr}%
  {\color{dynamodarkpurple}\emph{Case study~\themyctr: #1}}
}
\author{Jiahui Xu}
\affiliation{%
  \institution{ETH Zurich}
  \city{Zurich}
  \country{Switzerland}
}
\email{jxu@ethz.ch}

\author{Emmet Murphy}
\affiliation{%
  \institution{ETH Zurich}
  \city{Zurich}
  \country{Switzerland}
}
\email{emurphy@ethz.ch}

\author{Lana Josipovi\'c}
\affiliation{%
  \institution{ETH Zurich}
  \city{Zurich}
  \country{Switzerland}
}
\email{ljosipovic@ethz.ch}

\let\defaultautoref\autoref
\renewcommand*{\autoref}[2][]{%
  \hyperref[{#2}]{%
    \defaultautoref*{#2}%
    \ifx\\#1\\%
    \else
      #1%
    \fi
  }%
}

\begin{abstract}

When the MLIR project was first introduced, it promised to address the issues that the HLS community had with the LLVM project.
But is this really the case, and is MLIR the "right"/"best" compiler infrastructure for HLS? We here share our experiences based on the development of Dynamatic~(\href{https://github.com/EPFL-LAP/dynamatic}{github.com/EPFL-LAP/dynamatic}).

\end{abstract}

\begin{document}

%%
%% The "title" command has an optional parameter,
%% allowing the author to define a "short title" to be used in page headers.
\title[Is It a Good Idea to Build an HLS Tool on Top of MLIR?]{{Is It a Good Idea to Build an HLS Tool on Top of MLIR? Experience from Building the Dynamatic HLS Compiler}}

\makeatletter
\renewcommand\@titlefont{\huge\bfseries} % Change \Large to \small or other sizes
\makeatother

\maketitle
\pagestyle{empty}

\section{Introduction}

The LLVM project~\cite{LLVM} has been a foundation for many open-sourced and commercial \emph{high-level synthesis}~(HLS) projects~\cite{XilinxVivadoHLS,CanisSep13,DynamaticMlir,JosipovicFeb20tut,CatapultHLS,BambuHLS}, that compile high-level software code to RTL.
However, LLVM is not ideal for HLS tools~\cite{MLIRPaper}: the IR cannot be customized to represent circuits, which forces HLS tools to create custom and hard-to-reuse IRs for circuits. 
% \lana{i am not sure i fully get this sentence: if it is too low-level, why would you need custom circuit representations? what makes it too low-level? I wouldnt say that MLIR dialects are higher-level than LLVM IR...}
%
% Due to these drawbacks, HLS tools typically use LLVM's frontend~(Clang) and optimization passes to convert C to optimized LLVM IR, and convert the LLVM IR to a custom HLS IR to perform the remaining HLS flow.
%
% Having different custom representations means that we have a fragmented market for different HLS tools: these tools cannot integrate with each other, and different frameworks have to re-invent the wheel.

MLIR~\cite{MLIRPaper,MLIRWebsite} promises to solve this issue: it introduces a standard way to define, create, analyze, and transform custom IR operations.
With MLIR, HLS developers can overcome the rigidity of LLVM IR: they can define high-level IRs that benefit from high-level transformations or low-level IRs with circuit semantics. 
For custom IRs, MLIR provides some default C++ routines for creating and manipulating IR objects, dumping and parsing their textual format, and so on.
This greatly reduces implementation overhead %; as a basis for other compilers, it also 
and encourages interoperability between the MLIR-based tools.
%
% The Dynamatic HLS compiler is built on top of MLIR.
% % It attempts to take advantage of all these aspects.
% Dynamatic implements a custom \emph{handshake} MLIR dialect to represent dataflow units; these units operate on a custom \emph{channel} data type to distinguish from normal wires.
% %
% As some of the main developers of Dynamatic, we could like to share our experience of its development with the LATTE community.

%
As some of the main developers of Dynamatic~\cite{DynamaticMlir,JosipovicFeb20tut}---an MLIR-based HLS tool---we acknowledge and leverage its abovementioned benefits.
Yet, we have also recognized some features of it that create obstacles.
This paper shares with the LATTE community the issues we encountered with MLIR~(as of version \href{https://github.com/llvm/llvm-project/tree/d8eb4ac41d881a19bea7673d753ba92e6a11f5d6}{d8eb4ac}): the IR definition, analysis, and transformation, and integration between different MLIR-based HLS projects. 
We believe that our observations are general and go beyond HLS MLIR projects.
We hope that these findings will bring new insights and discussions in the MLIR community and help the developers to make better decisions for future HLS tools.
% \lana{generalize to other projects...}
% In this paper, we first provide the reader with an overview of the architecture of Dynamatic, and then discuss some generic issues or limitations of the core function of MLIR that might be problematic for HLS tools. 

% \jiahui{I don't think we need to write this much on LLVM, but in case we need to mention it, we can say some of the things below.}

% The LLVM project has been a fundamental building block of many open-source and commercial HLS tools. However, this is far from an ideal solution:
% %
% Extending the LLVM IR has always been a logistical nightmare: custom instructions and data types cannot be easily introduced without applying a significant modification to LLVM, which inevitably causes a significant divergence of the LLVM versions.
% %
% The LLVM IR is too low-level to apply high-level optimization: LLVM even introduces Polly to perform high-level loop nest optimization---which relies on a custom format that wraps around the LLVM IR.
% %
% The LLVM IR's semantics are also heavily influenced by CPU ISAs: for example, many arithmetic operations~(e.g., add and mul) require the same data types for the inputs and output, which is far from the convention that is typically used in custom hardware designs.
% %
% Due to these drawbacks, HLS tools typically reuse LLVM's C/C++ frontend and optimization passes, and convert the software IR to an ad-hoc IR to perform the remaining HLS steps.
% %

% \clearpage

\section{Background}
This section discusses the relevant background of MLIR and Dynamatic.

% \jiahui{TODO: Introduce HLS of dataflow circuit.}
% \jiahui{TODO: If space is a problem, do not mention the detailed list of transformation passes}

% \begin{figure}[h!]
%     \centering
%     \includegraphics[width=\linewidth]{figs/architecture.drawio.pdf}
%     \caption{The compilation pipeline of Dynamatic}
%     \label{fig:is-mlir-good:dynamatic-pipeline}
% \end{figure}

%

MLIR~\cite{MLIRWebsite} is a compiler infrastructure that helps define custom IR~(called dialects) and transformation passes.
The IR contains \emph{operations}, which consume and produce \emph{values}.
%
% Above the operations, MLIR defines two levels of hierarchy: a block contains a list of operations, and a region contains a list of blocks. An MLIR region can be an \emph{SSA region}: this means that the region describes a set of sequentially executed blocks~(i.e., basic blocks) and operations.
% A region can be a \emph{graph region}: this means that the operations in the region model a graph.
% An operation can also have a region within; for example, an MLIR function is an operation that contains a region.

Dynamatic produces \emph{dataflow circuits}~\cite{CortadellaJul10,JosipovicFeb18}, which consist of \emph{dataflow units} of instruction granularity connected via \emph{handshake channels}; the data is encapsulated in a token, exchanged via handshake channels.
In dataflow circuits, operations execute whenever their inputs are valid. Therefore, Dynamatic produces dynamically-scheduled circuits that have a performance advantage whenever the control flow or memory access pattern is unpredictable~\cite{JosipovicFeb19, JosipovicSep17, ElakhrasMar24}.
%
% \autoref{fig:is-mlir-good:dynamatic-pipeline} illustrates the compilation pipeline of Dynamatic, consisting of the following building blocks:
Dynamatic is an MLIR-based compiler: it represents units and channels as operations and values in the specialized \emph{handshake} dialect.
%since the operations do not execute in any particular order, the IR is contained in a graph region.

Dynamatic has evolved beyond a research prototype: It forms the basis for numerous publications~\cite{JosipovicFeb18,JosipovicSep17,JosipovicFeb19,JosipovicDec19,JosipovicFeb20,ChengFeb20,JosipovicMar21,JosipovicMay22,ChengAug22,ElakhrasAug22,RizziAug22,LiuDec22,ElakhrasFeb23,XuFeb23,RizziJul23,XuOct23,WangNov23,ElakhrasMar24,XuMar24,GuerrieriSep24,LiuSep24,XuApr25,ElakhrasApr25,BouilloudMay25,KatsumiFeb26,PirayadiFeb26,HerklotzMar26}---many of which have been incorporated into the main HLS flow~\cite{JosipovicFeb18,JosipovicSep17,JosipovicFeb19,JosipovicDec19,JosipovicFeb20,ElakhrasAug22,RizziAug22,LiuDec22,ElakhrasFeb23,XuFeb23,WangNov23,XuApr25}.
We have held multiple tutorials and talks in technical conferences~\cite{JosipovicFeb20tut,JosipovicMar24tut}. Dynamatic merges approximately 30 pull requests (PR) every month into its main branch, and every PR is monitored with an automated CI/CD pipeline to prevent compilation errors or performance regression. Code contributions from less experienced developers will receive detailed reviews and feedback.

% Within Dynamatic, we develop custom LLVM 
\begin{comment}

\jiahui{Maybe the compilation pipeline is not particularly interesting for the main points of the paper.}

%
Dynamatic relies on Clang and LLVM passes to convert C code to an optimized LLVM IR.
Within Dynamatic, we develop a custom LLVM pass~\cite{JosipovicDec19} to identify the dependencies between memory accesses. Dynamatic later instantiate specialized memory controller to ensure that the dependencies are always honored. %, which will be used to identify the array accesses that need to be connected using a load-store queue.
The LLVM IR is converted to the ControlFlow MLIR dialect to benefit from the MLIR infrastructure.
% (4)~We apply a set of software optimizations that are not available in LLVM or are insufficient for custom hardware compilation~(e.g., custom bitwidth optimizations). 
We translate the ControlFlow dialect to the handshake dialect circuit~\cite{JosipovicFeb18}.
We apply buffer insertion to regulate the critical path and optimize circuit throughput~\cite{JosipovicFeb20}. 
% during this step, we share scarce functional units and calculate the appropriate sizes for the load store queues. This step is conceptually similar to the scheduling, allocation, and binding in static-scheduling-based HLS tools.
We translate the handshake dialect to an RTL circuit.

\end{comment}

% The following discusses the limitations of MLIR on IR representation, transformation, and project integration.

\section{Building an MLIR-Based HLS Compiler \\ in an Academic Setting}
We acknowledge the significant advantages of using MLIR as the basis for an HLS project in an academic setting. Dynamatic is mainly built by student developers. Many of us have an electrical engineering background, and we are not professionally trained software developers; often, we are unaware of best software practices.
MLIR helps us overcome this challenge: it provides insight into how a complex object-oriented programming architecture works in practice, and its organization has inspired us to create a modular, easy-to-understand tool.
%
%The organization of the MLIR project has inspired us to create a modular, easy-to-understand tool.
%
These benefits enable us to dedicate more development effort to HLS-specific features.
% \lana{but there are also problems, as we discuss next}

% However, there are also issues, 
However, we also recognize that MLIR has features that create obstacles for HLS tools, as we discuss next.

% \clearpage

\section{Limitations of MLIR for Open-Source HLS Projects and Case Studies in Dynamatic}

This section discusses the issues that we encountered when building the Dynamatic HLS tool on top of MLIR.

% We would like to start by acknowledging the profound benefit of MLIR as a foundation for an HLS project in a university environment: Dynamatic is built mainly by student developers, many of us have an electrical engineering background and are not trained software developers, and often do not know good practices.

% \lana{Could you make the captions more provocative? Everything here "has an overhead", "is nontrivial", "is challenging", which does not make me worried--everything in life is challenging, so what... :P Try to go with some punchy lines or questions (Where are the circuit edges? Where did the phis go? What does it take to combine MLIR projects? I'm rambling a bit but you see what I mean...)
% }

% This section discusses the features of MLIR that might be inconvenient or problematic for building an HLS compiler; we present case studies obtained from our development experience.

% This section discusses the features of MLIR that Dynamatic failed to take advantage of.

\subsection{MLIR Value is Not a Fit for Modeling Graph Edges}
\label{sec:edge-info}

% This section describes the limitations from an IR representation perspective.

%
Modeling software IRs or circuits as graphs is a common practice~\cite{DeMicheli94}. 
Some information naturally belongs to the nodes, and some belongs to the edges.
MLIR allows annotations~(called \emph{attributes}) on operations.
%
% The mechanism is similar to attaching metadata to LLVM instructions.
%
% Unlike LLVM, where each value is associated with a unique operation, an MLIR operation can output multiple values.
However, no attribute can be attached to MLIR values, representing data exchanged between operations.
How does an MLIR HLS framework annotate edge information?
\casestudy{memory dependencies.} \autoref{fig:edge-info} describes a C program and a fraction of the corresponding MLIR. There is a potential read-after-write dependency between \texttt{store3} and \texttt{load2} from the following iteration. This dependency is typically determined by software IR analysis; a \emph{dependency edge} between the store and the load is annotated with the \emph{dependence distance} and used as an HLS scheduling constraint to guarantee correct operation order. As MLIR disallows edges or operation pairs to be annotated, this information must be represented in a non-standard and non-intuitive way.

Dynamatic uses alias and polyhedral analysis to determine the dependency pairs between memory accesses~\cite{JosipovicDec19,JosipovicSep17}.
To maintain this information in the HLS flow, Dynamatic assigns every operation a unique name, maintains the name throughout the HLS pipeline, and uses a hard-to-read format to represent the dependency.
For correctness, Dynamatic must additionally ensure the validity of the dependency by guaranteeing that every transformation preserves the pairwise validity of the names.

\casestudy{recording software profiling.}
% Profiling branches is important for HLS compilation---we want to focus on frequently executed loops to effectively shorten the overall latency. It is beneficial to represent them in the IR, as many optimization passes might be interested in them.
%
% Since transitions are described between pairs of basic blocks, it is natural to describe them as edge information---which also needs a nonstandard extension to MLIR.
Software profiling is typically used in HLS to determine operation execution counts and sequences (e.g., for early performance estimates or targeted optimizations of frequently executed constructs). Although this information is naturally attributed to control-flow edges between basic blocks, there is no standard way to annotate control-flow edges in MLIR.

% \lana{This sounds a bit too narrow/specific to our used case. Maybe it could be said in a more general way (and you narrow it down when you talk about Dynamatic): Software profiling is a typically used in HLS to determine operation execution counts and sequences (e.g., for early performance estimates or targeted optimizations of frequently executed constructs). Although this information is naturally attributed to control flow edges between basic blocks, there is no standard way to annotate control flow edges in MLIR. }

Dynamatic relies on software profiling to identify the frequently executed loops that should be prioritized in optimization~\cite{JosipovicFeb20,RizziAug22}.
This information could have been recorded on the output edges of the branch units~(see the \texttt{cond\_br} nodes in \autoref[b]{fig:dialect-conversion-limitation}).
% Since the basic block transition is defined on a pair of basic blocks, this information is naturally edge information.
Yet, since MLIR does not allow value annotation, Dynamatic had to rely on an external CSV file to hold this information.
% \lana{line break if you refer to both issues (cstudy1 and cstudy2?}

Proper edge-annotation support in MLIR would alleviate these issues.

% \lana{Here and everywhere: I know I suggested to discuss Dynamatic last, but I find the paragraphs somewhat unfinished/hanging. having some general concluding sentence ('a proper edge annotation support in MLIR would alleviate these issues' or whatever) would tie it all together nicely}

\begin{figure}[t!]
\footnotesize
\begin{subfigure}[t]{\linewidth}
\begin{Verbatim}[formatcom=\color{black},vspace=0pt]
void histogram(in_int_t feature[1000], in_float_t weight[1000],
               inout_float_t hist[1000], in_int_t n) {
  for (int i = 0; i < n; ++i) {
    int m = feature[i]; 
    float wt = weight[i];
\end{Verbatim}
\begin{Verbatim}[formatcom=\color{dynamopurple},vspace=0pt]
    float x = hist[m]; // -> "name = load2"
\end{Verbatim}
\begin{Verbatim}[formatcom=\color{dynamopurple},vspace=0pt]
    hist[m] = x + wt; // -> "name = store3"
\end{Verbatim}
\begin{Verbatim}[formatcom=\color{black},vspace=0pt]
  }
}
\end{Verbatim}
\caption{A program with a \emph{read-after-write}~(RAW) dependency.}
\end{subfigure}
\label{fig:placeholder}
\begin{subfigure}[t]{\linewidth}
\begin{Verbatim}[formatcom=\color{black},vspace=0pt]
%9 = memref.load %feature[%8] {
  handshake.name = "load2"
  } : memref<1000xf32> 
...
// The dependency information, which conceptually
// is an edge "store3 -> load2" is annotated 
// on the node in an awkward way.
memref.store %10, %feature[%11] {
\end{Verbatim}
\begin{Verbatim}[formatcom=\color{dynamopurple},vspace=0pt]
  handshake.deps = #handshake<deps[["load2", 1]]>,
\end{Verbatim}
\begin{Verbatim}[formatcom=\color{black},vspace=0pt]
  handshake.name = "store3"
  } : memref<1000xf32>
\end{Verbatim}
\caption{
The produced MLIR snippet.
%We can only store the edge information on the predecessor node. The format is not self-explanatory and creates a high entry barrier to the project.
}
\end{subfigure}
% \vspace{.5em}
\caption{
Annotating edge information in MLIR is tricky.
%A memory dependency describe the execution sequence requirement between \emph{a pair of nodes}---an information that is natually encoded as an edge. This is not natively achievable in MLIR.
}
\label{fig:edge-info}
\end{figure}

% \subsection{Does MLIR Help Represent a Dataflow Circuit?}

\subsection{Are Block Arguments Convenient \\ for HLS Conversion?}
\label{sec:block-args}
\begin{figure}[t!]
\footnotesize
% \centering
\begin{subfigure}[b]{.49\linewidth}%
% \begin{Verbatim}[formatcom=\color{black},vspace=0pt]
%     ^bb1(%3 : i32, %4 : i32):
%       %5 = addi %3, %2 : i32
%       %pred = cmpi slt %4, %5 : i1
%       cf.cond_br %pred, ^bb2(%5), ^bb3(%5)
% \end{Verbatim}
\begin{verbatim}
^bb1(%3 : i32, %4 : i32):
%5 = addi %3, %2 : i32
%pred = cmpi slt %4, %5 : i1
cf.cond_br %pred,^bb2(%5),^bb3(%5)
\end{verbatim}
% \subcaption{A basic block in the ControlFlow dialect}
\subcaption{}
\end{subfigure}%
~
\hfill%
\begin{subfigure}[b]{.49\linewidth}
\centering
\includegraphics[width=.8\linewidth]{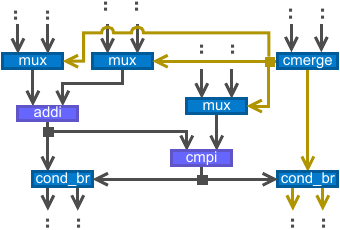}
% \subcaption{The corresponding dataflow circuit.}
\subcaption{}
\end{subfigure}
\caption{Dynamatic aims to translate software IR (left) to dataflow circuit (right).
% Dynamatic uses the rewrite infrastructure in MLIR to carry out the conversion.
% \jiahui{Need to say what rewrite can do first.}
% However, this appears to be a big challenge since (1)~there are no SSA Phi nodes in MLIR, so there is nothing to be \emph{rewritten} into a mux in the dataflow circuit; (2)~We need to flatten the SSA basic blocks into a graph region to allow cyclic paths; (3)~the basic data types~(e.g., \texttt{i32}) has to be converted into specific channel types~(e.g., \texttt{handshake.channel<i32>}), which requires type conversion.
}
\label{fig:dialect-conversion-limitation}
\end{figure}

% This section describes the limitations from a dialect transformation perspective.

% MLIR can \emph{represent} both software and circuits.
%
In an HLS flow, we need to convert a software representation into a circuit.
In particular, we have to convert SSA $\phi$ nodes~(representing a conditional assignment) into multiplexers.
Is this conversion straightforward in MLIR?
% This conversion is not straightforward in MLIR, as described next.
% Although MLIR provides useful utilities for the transformation within one region type, this HLS conversion is not straightforward, as described next.
%

%
\casestudy{convert block arguments to multiplexers.}
% However, MLIR implement the SSA $\phi$ nodes as the arguments of the blocks~(block arguments), which does not have any information attached about the incoming values.
\autoref[a]{fig:dialect-conversion-limitation} describes a basic block in the ControlFlow dialect~(a built-in MLIR dialect for representing sequential programs); it begins with an identifier~(bb1) with the \emph{block arguments}~(\texttt{\%3, \%4}) and ends with a conditional branch.
MLIR represents the $\phi$ nodes with these block arguments~\cite{MLIRBlockArgumentRationale}. This representation has the following issues for the software to circuit transformation:
(a)~The block arguments are \emph{values} with no producer.
(b)~The branches collect the outgoing values, but these values are disconnected from the successor blocks.
The former issue makes pattern rewriting---the standard method for converting between dialects---a poor fit for this conversion, as there is no operation to match. The latter issue makes locating the inputs to the multiplexers unnecessarily complex: unlike an LLVM $\phi$ node, where the input values from the other BBs are directly accessible, to obtain the same information, we need to first locate the parent BB, the branch, and then the values fed into the branch.

% These features create some challenges for a clean conversion:
% %
% Since block arguments and branches are syntactically very different from multiplexers~(c.f., the dataflow circuit in \autoref{fig:dialect-conversion-limitation}b vs. the MLIR SSA region in \autoref{fig:dialect-conversion-limitation}a), this transformation cannot be benefit from local pattern rewriting~(there is no subgraph to be rewritten from).
% %
% Locating the inputs for the select is intrinsically a convoluted process:
% unlike an SSA $\phi$ node in LLVM where the incoming values are directly accessible, to obtain the same information, we need to first locate the parent BB, the branch, and then the values fed into the branch.
% \autoref{fig:dialect-conversion-limitation} describes an Dynamatic example of converting from the ControlFlow dialect~(a built-in dialect for representing SSA IR) to dataflow circuit.

Dynamatic uses MLIR's pattern rewriting to carry out this conversion.
We acknowledge that \emph{this solution is not ideal} since the rewrite must operate on the entire function~(which defeats the purpose of the local rewrite rules).
% (2)~Dynamatic uses specialized data types for representing channels~(e.g., \texttt{channel<i32>}), while the rewrite utility in MLIR becomes very complex to use when type conversion is involved.

%

\begin{comment}

%
Pattern rewriting is a widely used utility for optimizations or dialect conversion.
%
Developers typically specify a \emph{match-and-rewrite} function that matches a \emph{root operation} of a specific type, visiting its neighboring operation to see if it matches a certain subgraph pattern, and replacing a subgraph with a newly created subgraph.
%
However, this utility is not very helpful for conversion between dialects that have a large syntactic difference.

%
%
In any HLS strategy, condition assignments will eventually be materialized as multiplexers. 
Instead of SSA $\phi$ nodes, MLIR uses block arguments to represent conditional assignments.
%
However, block arguments are not operations and cannot be matched by any pattern; therefore, it is impossible to implement this conversion as a (local) rewrite pattern.
%

\end{comment}

%
% This conversion has to map the block arguments~(\texttt{bb1(\%3, \%4)}) back to $\phi$ nodes~(implemented as \texttt{mux}), which could not take advantage of the rewrite transformation utility provided by MLIR.

% \subsection{Can We Annotate IR Information Nicely?}

\subsection{Have We Solved the Software Segmentation Problem?}
\label{sec:sw-seg}

% This section describes the limitations from a project integration perspective.

%
MLIR promises to resolve the software fragmentation problem by allowing the compiler projects to reuse each other's transformation passes.
This might be true within the official MLIR repository, as the maintainers have a coherent view on how the segments would work together.
However, many projects are not inside the upstream MLIR project, for example, CIRCT~\cite{CIRCT}, Polygeist~\cite{PolygeistPaper}, Dynamatic~\cite{DynamaticMlir}, MLIR-AIE~\cite{MLIRAIE}, etc.
%
% There is a logistical challenge when integrating these projects.
Can we smoothly integrate these projects?
% The code reuse and integration challenges between different external MLIR-based projects still exist.
%

\begin{figure}[t!]
\centering
\includegraphics[width=0.8\linewidth]{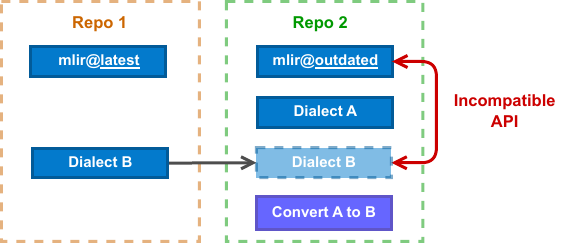}
\caption{{Integrating two MLIR projects is challenging.}}
\label{fig:fragmentation-problem}
\end{figure}

\casestudy{implement a transformation pass for dialects maintained in two GitHub repositories.}
\autoref{fig:fragmentation-problem} describes simplified architectures of two MLIR-based projects.
These projects implement two different dialects, A and B. Suppose that we want to implement a new MLIR pass that converts dialect A to B. 
It is conceptually infeasible if the implementations of dialects A and B are dependent on two different LLVM projects with mismatching APIs~(compilation error in either MLIR versions).

Dynamatic implements a custom translation pass that converts the handshake dialect into the XLS MLIR dialect~\cite{GoogleXls}.
Since it is impractical for us to constantly update and synchronize Dynamatic's LLVM version with XLS~(which is updated daily), this feature has a risk of getting outdated very quickly.

\subsection{Do We Have a Good HLS C Frontend \\that Encourages People to Build New HLS Tools?}
\label{sec:c-front}
An HLS \emph{frontend} converts C code to IR, applies IR optimizations, and annotates important information for circuit generation.
Is there a C frontend for MLIR that satisfies all these requirements?

% \casestudy{Who provides IR optimizations?}
% % \lana{what are THESE opts?}
% %
% % Since LLVM provides a set of powerful and well-tested IR optimizations, most MLIR-based projects treat LLVM IR as the final target of their transformation, and thus \emph{no one is incentivized to implement a translation pass from LLVM IR to MLIR}.
% Since LLVM provides a set of powerful and well-tested IR optimizations, most MLIR-based projects are still reliant on LLVM IR. Therefore, MLIR-based projects are not incentivized to implement these optimization passes.
% To be a good fit for MLIR-based HLS tools, MLIR must replace the LLVM IR.
% %, as an optimized IR is the starting point of an HLS flow.
% % This is an issue for the HLS community: unlike other projects, an optimized IR is the starting point of an HLS flow.

% % Thus, any flow must go through LLVM IR to benefit from these optimizations without reimplementing all of them.
% % Due to this reason, although there is an MLIR LLVM dialect that developers can translate LLVM IR into, there is no incentive in the community to implement the translation from this IR to some other dialects. 

% \casestudy{The status of C-to-MLIR conversion.} 
% %
% Some projects convert C to MLIR: \emph{Polygeist} converts C AST to MLIR SCF dialect; the \emph{CIR project} in Clang converts C AST to a CIR dialect~(a dialect that models the C semantics).
% %
% These solutions alone do not provide fine-grained IR optimizations. To use them as the C frontend, we inevitably need to go through LLVM IR---which MLIR \emph{has aimed to avoid in the first place}.

\casestudy{The status of C-to-MLIR conversion.} Some projects convert C to MLIR: \emph{Polygeist} converts C AST to MLIR SCF dialect; the \emph{CIR project} in Clang converts C AST to a CIR dialect~(a dialect that models the C semantics). These solutions alone do not provide fine-grained IR optimizations, making them non-competitive with LLVM-based HLS compiler frontends. 

\casestudy{The reliance on LLVM.} Since LLVM provides a set of powerful and well-tested IR optimizations, most MLIR-based projects eventually lower their custom dialect to LLVM IR to benefit from LLVM IR optimizations. The ability to exploit the LLVM IR removes the incentive to implement the same optimizations in MLIR. Unfortunately, it also introduces all LLVM limitations into the compiler pipeline that MLIR aimed to avoid in the first place. 

Since the existing MLIR C front-ends are not competitive with LLVM-based ones, Dynamatic is therefore still reliant on the LLVM IR for this process.
We have to use custom workarounds for performing memory analysis and infer the array sizes from the original C code---these workarounds, difficult to implement in the rigid LLVM framework, would fit naturally in the MLIR compiler flow.
% \lana{why come back at the end of the paragraph to the same point that you started the paragraph with? You could say: Since the existing MLIR C front-ends are not competitive with LLVM-based ones, Dynamatic uses Clang... ... original C code.} 
% \lana{Consider ending with something that emphasizes the problem of the custom workarounds like: we have to use custom workarounds for performing memory analysis and infer the array sizes from the original C code---these workarounds, difficult to implement in the rigid LLVM framework, would fit naturally in the MLIR compiler flow. [or whatever is your point in saying that--if the issues would be the same in a proper MLIR frontend, then I am not sure they are worth bringing up?]}

% This solution is \emph{not ideal}, as it requires lifting the abstraction level: LLVM IR does not record type information of pointers in function arguments. This includes the dimensions of the fixed-sized arrays, which are used to determine the index widths used in the RTL. Dynamatic reuses the original C code to recover this information. Due to non-trivial conversions like this, it is impractical for us to develop a fully general solution for HLS tools.

% \section{Benefits of MLIR}
\section{Beyond HLS}
% \lana{Added this section to address the reviews, pls check}

Although we have here focused on our particular use case and HLS experiences, we believe that the issues above are general and go beyond just HLS MLIR projects: 
(1)~Both software and hardware compilers routinely exploit edge-specific information (e.g., memory dependencies, branch probabilities) and may suffer MLIR's limited expressiveness discussed in \autoref{sec:edge-info}. 
(2) The SSA challenge of \autoref{sec:block-args} is primarily hardware-oriented; yet, the lack of $\phi$s may complicate IR optimizations and analyses based on use-def chains and data dependencies in software compilers as well. 
(3) The software segmentation problem of \autoref{sec:sw-seg} is perfectly general and applies to any MLIR project. 
(4) Any compiler flow that uses C as the input language suffers from the limitations discussed in \autoref{sec:c-front}.

We therefore believe that addressing these concerns would benefit the general MLIR community. 

\section{Conclusion}

We discussed some major features of MLIR that lead to fragile workarounds in Dynamatic.
We look forward to discussing the lessons learned and the future development of MLIR-based HLS projects with the LATTE community.

% \subsection{Does MLIR Help Us Implement Transformation/Analysis Pass}

% \subsection{Does MLIR Solve the Software Fragmentation Problem?}

% \section{Lessons Learned: What We Got Wrong about MLIR}

\clearpage
\balance
% \bibliographystyle{ACM-Reference-Format}
% \bibliography{{./sample-base.bib}}

\clearpage

\printbibliography

@string{CASES = "Proceedings of the International Conference on
Compilers, Architectures, and Synthesis for
Embedded Systems"}

@string{FPGA18  = "Proceedings of the 26th {ACM}/{SIGDA} International
Symposium on Field Programmable Gate Arrays"}

@string{FPGA19  = "Proceedings of the 27th {ACM}/{SIGDA} International
Symposium on Field Programmable Gate Arrays"}

@string{FPGA20  = "Proceedings of the 28th {ACM}/{SIGDA} International
Symposium on Field Programmable Gate Arrays"}

@string{FPGA23  = "Proceedings of the 31st {ACM}/{SIGDA} International
Symposium on Field Programmable Gate Arrays"}

@string{FPGA24  = "Proceedings of the 32nd {ACM}/{SIGDA} International
Symposium on Field Programmable Gate Arrays"}

@string{FPL22  = "Proceedings of the 22nd International Conference on
Field-Programmable Logic and Applications"}

@string{FPT  = "Proceedings of the {IEEE} International Conference
on Field Programmable Technology"}

@string{ICCAD23   = "Proceedings of the 42nd International Conference on
Computer-Aided Design"}

@string{MEMOCODE10 = "Proceedings of the 10th {ACM}/{IEEE} International 
Conference on Formal Methods and Models for Codesign"}

@string{TECS    = "{ACM} Transactions on Embedded Computing Systems"}

@InProceedings{ElakhrasFeb23,
author =   {Elakhras, Ayatallah and Sawhney, Riya and Guerrieri, Andrea and Josipovi\'{c}, Lana and Ienne, Paolo},
title =  {Straight to the Queue: Fast Load-Store Queue Allocation in Dataflow Circuits},
booktitle = {Proceedings of the 31st International Symposium on Field-Programmable Gate Arrays},
address =  {Monterey, CA},
month =  feb,
year =   2023,
pages =  {39--45},
doi = {10.1145/3543622.3573050},
url = {https://doi.org/10.1145/3543622.3573050}
}

@InProceedings{ElakhrasMar24,
author =   {Elakhras, Ayatallah and Guerrieri, Andrea and Josipovi\'{c}, Lana and Ienne, Paolo},
title =  {Survival of the Fastest: Enabling More Out-of-Order Execution in Dataflow Circuits},
booktitle = {Proceedings of the 32nd International Symposium on Field-Programmable Gate Arrays},
address =  {Monterey, CA},
month =  mar,
year =   2024,
pages =  {44--54},
doi = {10.1145/3626202.3637556},
url = {https://doi.org/10.1145/3626202.3637556},
}

@InProceedings{ElakhrasAug22,
author = {Elakhras, Ayatallah and Guerrieri, Andrea and Josipovi\'{c}, Lana and Ienne, Paolo},
title = {Unleashing Parallelism in Elastic Circuits with Faster Token Delivery},
booktitle = {Proceedings of the 32nd International Conference on Field-Programmable Logic and Applications},
address = {Belfast, UK},
month = aug,
year = 2022,
pages = {253--61},
doi = {10.1109/FPL57034.2022.00046},
url = {https://doi.org/10.1109/FPL57034.2022.00046},
}

@InProceedings{JosipovicFeb18,
author =   {Josipovi\'{c}, Lana and Ghosal, Radhika and Ienne, Paolo},
title =  {Dynamically Scheduled High-level Synthesis},
booktitle =  FPGA18,
address =  {Monterey, CA},
month =  feb,
year =   2018,
pages =  {127--36},
doi = {10.1145/3174243.3174264},
url = {https://doi.org/10.1145/3174243.3174264},
}

@InProceedings{JosipovicFeb20,
author =   {Josipovi\'{c}, Lana and Sheikhha, Shabnam and Guerrieri, Andrea and Ienne, Paolo and Cortadella, Jordi},
title =  {Buffer Placement and Sizing for High-Performance Dataflow Circuits},
booktitle =  FPGA20,
address =  {Seaside, CA},
month =  feb,
year =   2020,
pages =  {186--96},
doi = {10.1145/3373087.3375314},
url = {https://doi.org/10.1145/3373087.3375314}
}

@InProceedings{RizziAug22,
author = {Rizzi, Carmine and Guerrieri, Andrea and Ienne, Paolo and Josipovi\'{c}, Lana},
title =	 {A Comprehensive Timing Model for Accurate Frequency Tuning in Dataflow Circuits},
booktitle = FPL22,
address = {Belfast, UK},
month =	 aug,
year = 2022,
pages = {375--83},
doi = {10.1109/FPL57034.2022.00063},
}

@InProceedings{JosipovicMay22,
author =   {Josipovi\'{c}, Lana and Marmet, Axel and Guerrieri, Andrea and Ienne, Paolo},
title =  {Resource Sharing in Dataflow Circuits},
booktitle =  {Proceedings of the 30th {IEEE} Symposium on Field-Programmable Custom Computing Machines},
address =  {New York},
month =  may,
year =   2022,
pages =  {1--9},
doi = {10.1109/FCCM53951.2022.9786084},
url = {https://doi.org/10.1109/FCCM53951.2022.9786084},
}

@inproceedings{JosipovicFeb20tut,
author = {Lana Josipovi{\'{c}} and Andrea Guerrieri and Paolo Ienne},
title =	{{Dynamatic}: From {C}/{C}++ to Dynamically Scheduled Circuits},
booktitle =	 FPGA20,
address = {Seaside, CA},
month =	 feb,
year =	 2020,
pages =	 {1--10},
doi = {10.1145/3373087.3375391},
url = {https://doi.org/10.1145/3373087.3375391},
}

@Article{JosipovicMar21,
key =		 {Jos},
author =	 {Lana Josipovi{\'{c}} and Andrea Guerrieri and Paolo
                          Ienne},
title =	 {Synthesizing General-Purpose Code into Dynamically
            Scheduled Circuits},
journal =	 {{IEEE} Circuits and Systems Magazine},
year =	 2021,
volume =	 21,
number =	 2,
pages =	 {97--118},
month =	 {Second quarter},
doi =		 {10.1109/MCAS.2021.3071631}
}

@InProceedings{ChengFeb20,
author =   {Cheng, Jianyi and Josipovi\'{c}, Lana and Constantinides,
                                         George A. and Ienne, Paolo and Wickerson, John},
title =  {Combining Dynamic \& Static Scheduling in High-Level Synthesis},
booktitle =  FPGA20,
address =  {Seaside, CA},
month =  feb,
year =   2020,
pages =  {288--98},
doi = {10.1145/3373087.3375297},
}

@InProceedings{CortadellaJul10,
author =   {Cortadella, Jordi and Galceran-Oms, Marc and
            Kishinevsky, Mike},
title =  {Elastic systems},
booktitle =  MEMOCODE10,
month =  jul,
year =   2010,
pages =  {149--58},
}

@Article{JosipovicSep17,
key =    {Jos},
author =   {Lana Josipovi{\'{c}} and Philip Brisk and Paolo
                          Ienne},
title =  {An Out-of-Order Load-Store Queue for Spatial
          Computing},
journal =  TECS,
month =  sep,
year =   2017,
volume =   16,
number =   {5s},
pages =  {125:1--125:19},
doi =    {10.1145/3126525},
acceptance =   {\textbf{CASES'17 Best Paper Award Nominee}},
url = {https://doi.org/10.1145/3126525}
}

@InProceedings{JosipovicDec19,
author    = {Lana Josipovi\'{c} and Atri Bhattacharyya and Andrea Guerrieri and Paolo Ienne},
title     = {Shrink It or Shed It! {M}inimize the Use of {LSQ}s in Dataflow Designs},
booktitle = FPT,
year      = {2019}, 
month = dec, 
address = {Tianjin, China},
pages = {197--205},
url = {https://doi.org/10.1109/ICFPT47387.2019.00031}
}

@InProceedings{ChengAug22,
author = {Cheng, Jianyi and Josipovi\'{c}, Lana and Constantinides, George A. and Wickerson, John},
title = {Dynamic Inter-Block Scheduling for {HLS}},
booktitle = {Proceedings of the 32nd International Conference on Field-Programmable Logic and Applications},
address = {Belfast, UK},
month = aug,
year = 2022,
pages =  {243--52},
doi = {10.1109/FPL57034.2022.00045},
}

@Article{CanisSep13,
author =   {Canis, Andrew and Choi, Jongsok and Aldham, Mark and
            Zhang, Victor and Kammoona, Ahmed and Czajkowski,
            Tomasz and Brown, Stephen D. and Anderson, Jason H.},
title =  {{LegUp}: {A}n Open-Source High-Level Synthesis Tool
           for {FPGA}-Based Processor/Accelerator Systems},
journal =  TECS,
volume =   13,
number =   2,
month =  sep,
year =   2013,
pages =  {24:1--24:27},
address =  {New York},
doi={10.1145/2514740},
}

@InProceedings{JosipovicFeb19,
author =   {Josipovi\'{c}, Lana and Guerrieri, Andrea and Ienne, Paolo},
title =  {Speculative Dataflow Circuits},
booktitle =  FPGA19,
address =  {Seaside, CA},
month =  feb,
year =   2019,
pages =  {162--71},
doi = {10.1145/3289602.3293914},
}

@Manual{XilinxVivadoHLS,
organization = {Xilinx Inc.},
title =  {Vivado High-Level Synthesis},
url =    {http://www.xilinx.com/products/
          design-tools/vivado/integration/esl-design.html},
year =   2018,
key =   {VivadoHLS}
}

@Manual{LLVM,
organization = {The LLVM Compiler Infrastructure},
title =  {http://www.llvm.org},
url =    {http://www.llvm.org},
year =   2018,
key =    {LLVM}
}

@Manual{CIRCT,
organization = {CIRCT IR Compiler and Tools},
title =  {https://github.com/llvm/circt},
url =    {https://github.com/llvm/circt},
year =   2020,
key =    {CIRCT}
}

@Book{DeMicheli94,
key =    "Dem",
author =   "De Micheli, Giovanni",
title =  "Synthesis and Optimization of Digital Circuits",
publisher =  "McGraw-Hill",
address =  "New York",
year =   1994,
}

@inproceedings{XuFeb23,
author={Xu, Jiahui and Murphy, Emmet and Cortadella, Jordi and Josipovi\'{c}, Lana},
title={Eliminating Excessive Dynamism of Dataflow Circuits Using Model Checking},
booktitle=FPGA23,
address={Monterey, CA},
year={2023},
pages={27--37},
month=feb,
doi={10.1145/3543622.3573196},
url={https://doi.org/10.1145/3543622.3573196},
}

@inproceedings{XuOct23,
author={Xu, Jiahui and Josipović, Lana},
title={Automatic Inductive Invariant Generation for Scalable Dataflow Circuit Verification},
booktitle=ICCAD23,
year = {2023},
month = oct,
address = {San Francisco, CA},
pages={1--9},
doi={10.1109/ICCAD57390.2023.10323796},
url={https://doi.org/10.1109/ICCAD57390.2023.10323796},
}

@inproceedings{XuMar24,
author={Xu, Jiahui and Josipović, Lana},
title={Suppressing Spurious Dynamism of Dataflow Circuits Via Latency and Occupancy Balancing},
booktitle=FPGA24,
year = {2024},
month = mar,
address = {Monterey, CA},
pages = {188--98},
doi = {10.1145/3626202.3637570},
url = {https://doi.org/10.1145/3626202.3637570},
}

@manual{DynamaticMlir,
organization = {EPFL-LAP},
title = {{Dynamatic}}, 
url = {https://github.com/EPFL-LAP/dynamatic/tree/999dc3ce2fb95eac1dd39cad441fbdf6b8389aee},
}

@manual{GoogleXls,
title = {XLS: Accelerated HW Synthesis}, 
organization = {Google, Inc.},
url = {https://github.com/google/xls},
}

@inproceedings{BambuHLS,
author={Ferrandi, Fabrizio and Castellana, Vito Giovanni and Curzel, Serena and Fezzardi, Pietro and Fiorito, Michele and Lattuada, Marco and Minutoli, Marco and Pilato, Christian and Tumeo, Antonino},
booktitle={Proceedings of the 58th {ACM/IEEE} Design Automation Conference}, 
title={Invited: {Bambu}: an Open-Source Research Framework for the High-Level Synthesis of Complex Applications}, 
year={2021},
pages={1327--1330},
location={San Francisco, CA},
url={https://doi.org/10.1109/DAC18074.2021.9586110},
}

@inproceedings{MLIRPaper,
author={Lattner, Chris and Amini, Mehdi and Bondhugula, Uday and Cohen, Albert and Davis, Andy and Pienaar, Jacques and Riddle, River and Shpeisman, Tatiana and Vasilache, Nicolas and Zinenko, Oleksandr},
booktitle={Proceedings of the 19th {IEEE/ACM} International Symposium on Code Generation and Optimization}, 
title={{MLIR}: Scaling Compiler Infrastructure for Domain Specific Computation}, 
year={2021},
pages={2--14},
location={Seoul, Korea},
url={https://doi.org/10.1109/CGO51591.2021.9370308},
}

@manual{MLIRWebsite,
organization = {{LLVM} {Project}},
title =  {Multi-Level IR Compiler Framework ({MLIR})},
url = {https://mlir.llvm.org/},
year = 2020,
}

@manual{CatapultHLS,
organization={Siemens {EDA}},
title={{Catapult} High-Level Synthesis and Verification},
year=2026,
url={https://eda.sw.siemens.com/en-US/ic/catapult-high-level-synthesis/},
}

@manual{MLIRBlockArgumentRationale,
organization={{LLVM} Project},
title={{MLIR} Rationale: Block Arguments vs {PHI} nodes},
year=2026,
url={https://mlir.llvm.org/docs/Rationale/Rationale/#block-arguments-vs-phi-nodes},
}

@manual{MLIRAIE,
organization={AMD},
title={IRON API and MLIR-based AI Engine Toolchain
},
year=2026,
url={https://github.com/Xilinx/mlir-aie},
}

@inproceedings{PolygeistPaper,
author={Moses, William S. and Chelini, Lorenzo and Zhao, Ruizhe and Zinenko, Oleksandr},
booktitle={Proceedings of the 30th International Conference on Parallel Architectures and Compilation Techniques}, 
title={{Polygeist}: Raising C to Polyhedral {MLIR}}, 
year={2021},
pages={45--59},
doi={10.1109/PACT52795.2021.00011},
location={Atlanta, GA},
}

@inproceedings{LiuDec22,
  author    = {Liu, Jiantao and Rizzi, Carmine and Josipovi{\'c}, Lana},
  title     = {Load-Store Queue Sizing for Efficient Dataflow Circuits},
  booktitle = {Proceedings of the 21st International Conference on Field-Programmable Technology},
  pages     = {1--9},
  address   = {Hong Kong},
  month     = dec,
  year      = {2022},
  url={https://doi.org/10.1109/ICFPT56656.2022.9974425}
}

@inproceedings{RizziJul23,
  author    = {Rizzi, Carmine and Guerrieri, Andrea and Josipovi{\'c}, Lana},
  title     = {An Iterative Method for Mapping-Aware Frequency Regulation in Dataflow Circuits},
  booktitle = {Proceedings of the 60th Design Automation Conference},
  pages     = {1--6},
  address   = {San Francisco, CA},
  month     = jul,
  year      = {2023},
  url = {https://doi.org/10.1109/DAC56929.2023.10247686}
}

@inproceedings{WangNov23,
  author    = {Wang, Hanyu and Rizzi, Carmine and Josipovi{\'c}, Lana},
  title     = {{MapBuf}: Simultaneous Technology Mapping and Buffer Insertion for HLS Performance Optimization},
  booktitle = ICCAD23,
  pages     = {1--9},
  address   = {San Francisco, CA},
  month     = nov,
  year      = {2023},
  url = {https://doi.org/10.1109/ICCAD57390.2023.10323639}
}

@inproceedings{GuerrieriSep24,
  author    = {Guerrieri, Andrea and Guha, Srijeet and Lavin, Chris and Hung, Eddie and Josipovi{\'c}, Lana and Ienne, Paolo},
  title     = {{DynaRapid}: Fast-Tracking from C to Routed Circuits},
  booktitle = {Proceedings of the 34th International Conference on Field-Programmable Logic and Applications},
  pages     = {24--32},
  address   = {Turin, Italy},
  month     = sep,
  year      = {2024},
  url = {https://doi.org/10.1109/FPL64840.2024.00014}
}

@inproceedings{LiuSep24,
  author    = {Liu, Jiantao and Graczyk, Maksymilian and Guerrieri, Andrea and Josipovi{\'c}, Lana},
  title     = {Fast Switching Activity Estimation for HLS-Produced Dataflow Circuits},
  booktitle = {Proceedings of the 34th International Conference on Field-Programmable Logic and Applications},
  pages     = {118--125},
  address   = {Turin, Italy},
  month     = sep,
  year      = {2024},
  url = {https://doi.org/10.1109/FPL64840.2024.00025}
}

@inproceedings{XuApr25,
  author    = {Xu, Jiahui and Josipovi{\'c}, Lana},
  title     = {{CRUSH}: A Credit-Based Approach for Functional Unit Sharing in Dynamically Scheduled HLS},
  booktitle = {Proceedings of the 30th International Conference on Architectural Support for Programming Languages and Operating Systems},
  pages     = {249--263},
  address   = {Rotterdam, The Netherlands},
  month     = apr,
  year      = {2025},
  url = {https://doi.org/10.1145/3669940.3707273}
}

@inproceedings{ElakhrasApr25,
  author    = {Elakhras, Ayatallah and Xu, Jiahui and Erhart, Martin and Ienne, Paolo and Josipovi{\'c}, Lana},
  title     = {{ElasticMiter}: Formally Verified Dataflow Circuit Rewrites},
  booktitle = {Proceedings of the 30th International Conference on Architectural Support for Programming Languages and Operating Systems},
  pages     = {293--308},
  address   = {Rotterdam, The Netherlands},
  month     = apr,
  year      = {2025},
  url = {https://doi.org/10.1145/3676641.3715993}
}

@inproceedings{BouilloudMay25,
  author    = {Bouilloud, Mathias and Josipovi{\'c}, Lana and Luk, Wayne},
  title     = {Resource and Phase Awareness for Dynamically Scheduled High-Level Synthesis},
  booktitle = {Proceedings of the 15th International Symposium on Highly-Efficient Accelerators and Reconfigurable Technologies},
  pages     = {14--24},
  address   = {Kumamoto, Japan},
  month     = may,
  year      = {2025},
  url = {https://doi.org/10.1145/3728179.3728194}
}

@inproceedings{KatsumiFeb26,
  author    = {Katsumi, Shun and Murphy, Emmet and Josipovi{\'c}, Lana},
  title     = {{EagerlyElastic}: Correct-by-Construction Eager Execution in Dynamically-Scheduled HLS},
  booktitle = {Proceedings of the 34th International Symposium on Field-Programmable Gate Arrays},
  address   = {Seaside, CA},
  month     = feb,
  year      = {2026},
  pages = {103--113},
  url = {https://doi.org/10.1145/3748173.3779196}
}

@inproceedings{HerklotzMar26,
  author    = {Herklotz, Yann and Elakhras, Ayatallah and Camaioni, Martina and Ienne, Paolo and Josipovi{\'c}, Lana and Bourgeat, Thomas},
  title     = {{Graphiti}: Formally Verified Out-of-Order Execution in Dataflow Circuits},
  booktitle = {Proceedings of the 31st International Conference on Architectural Support for Programming Languages and Operating Systems},
  address   = {Pittsburgh},
  month     = mar,
  year      = {2026},
  note      = {To appear}
}

@misc{JosipovicMar24tut,
  author    = {Josipovi{\'c}, Lana and Ienne, Paolo and Ramirez, Lucas and Guerrieri, Andrea},
  title     = {{Dynamatic Reloaded}: An {MLIR}-Based Dynamically Scheduled HLS Compiler},
  url          = {https://www.isfpga.org/past/fpga2024/workshops-tutorials/},
  month     = mar,
  year      = {2024},
  note = {Tutorial at FPGA '24},
}

@inproceedings{PirayadiFeb26,
author = {Pirayadi, Rouzbeh and Elakhras, Ayatallah and Stojilovi\'{c}, Mirjana and Ienne, Paolo},
title = {Out with {LSQ}s: Custom Circuits for Memory Access Reordering in Dynamic HLS},
year = {2026},
url = {https://doi.org/10.1145/3748173.3779204},
doi = {10.1145/3748173.3779204},
booktitle = {Proceedings of the 34th International Symposium on Field-Programmable Gate Arrays},
pages = {92--102},
location = {Seaside, CA},
month = feb,
}

\end{document}
\endinput
%%
%% End of file `sample-sigconf.tex'.